\documentclass{iau} 
\usepackage{graphicx}
\usepackage{natbib}

\title[Common envelope: progress and transients] %% give here short title %%
{Common envelope: progress and transients}

\author[Natalia Ivanova]   %% give here short author list %%
{Natalia Ivanova$^1$}

\affiliation{$^1$Department of Physics, University of Alberta, Edmonton AB T6G 2E1
Canada \\ email: {\tt nata.ivanova@ualberta.ca}}

\pubyear{2016}
\volume{329}  %% insert here IAU Symposium No.
\jname{The Lives and Death--Throes of Massive Stars}
\editors{J.J. Eldridge, J.C. Bray, L.A.S. McClellandon \\ \& L. Xiao, eds.}
\doi{10.1017/S1743921317003398}
\begin{document}

\maketitle

\begin{abstract}
We   review  the   fundamentals   and  the   recent  developments   in
understanding of  common envelope physics.  We  report specifically on
the progress that  was made by the consideration  of the recombination
energy.   This energy  is found  to  be responsible  for the  complete
envelope ejection  in the case of  a prompt  binary formation,  for the
delayed dynamical ejections  in the case of  a self-regulated spiral-in,
and  for  the steady  recombination  outflows  during the 
transition between the plunge-in and the self-regulated spiral-in.  Due to
different  ways  how the  recombination  affects  the common  envelope
during  fast  and slow  spiral-ins,  the  apparent efficiency  of  the
orbital  energy  use  can  be  different  between  the  two  types  of
spiral-ins  by a  factor of  ten.  We  also discuss  the observational
signatures of  the common envelope events,  their link a new  class of
astronomical transients, Luminous Red Novae,  and to a plausible class
of very luminous irregular variables.

\end{abstract}
 
\firstsection % if your document starts with a section,
\section{Introduction: the energy sources and the energy sinks}

Common-envelope events (CEEs) are  fate-defining episodes in the lives
of close  binary systems.   During a common  envelope  phase, the
outer layers of  one of the stars expand to  engulf the companion, and
two stars  start to  temporarily orbit  within their  shared envelope.
This  pivotal  binary  changeover  ends with  a  luminosity  outburst,
leaving  behind either  a  significantly shrunk  binary,  or a  single
merged  star.   These  episodes  are  believed to  be  vital  for  the
formation  of  a  wide  range  of  extremely  important  astrophysical
objects,  including X-ray  binaries, close  double-neutron stars,  the
potential progenitors of Type Ia  supernovae and gamma-ray bursts, and
double black  holes that could produce  gravitational waves \citep[for
  more details on  overall importance of the CEEs, as  well as on many
  aspects of the involved physics, see the review in][]{CEreview}.

 The outcomes of the CEEs are believed to fall into two main divergent
 categories -- either a close binary formation, or a merger of the two
 stars  into a  single star.   The  boundary between  the outcomes  is
 usually found  by comparing the  available energy source  (the energy
 difference between  the orbital  energies before  and after  the CEE,
 $\Delta E_{\rm  orb}$), and the  required energy expense  (the energy
 required to displace the envelope to infinity, $E_{\rm bind}$).  This
 is        known         as        the         energy        formalism
 \citep{1984ApJ...277..355W,1988ApJ...329..764L}:

\begin{equation}
  \alpha \Delta E_{\rm orb}  = E_{\rm bind}= \frac{Gm m_{\rm env}}{\lambda R}
  \label{eq:aform}
\end{equation}

\noindent Here, $\alpha$  is the efficiency of the use  of the orbital
energy, and  it can be  only less  than one.  $m$  is the mass  of the
donor star  -- the star  whose expanded  envelope has formed  the CE,
$m_{\rm env}$ is the  mass of that envelope, $R$ is  the radius of the
donor. The parameter $\lambda$  relates  the envelope's binding energy $E_{\rm bind}$
as integrated from the stellar structure  with its parameterized form.

\newpage

This   famous   equation,   while   seems  to   be   transparent   and
straightforward, buries  a lot  of not  yet fully  understood physics.
For example,  there are  plenty of uncertainties  in how  to determine
$E_{\rm bind}$. That  includes such questions as what  is the boundary
between the core and the  ejected envelope, whether the thermal energy
can be converted effectively in the mechanical energy of the envelope,
and whether the out-flowing envelope  should be evaluated using
Bernoulli integral which is inclusive of the $P/\rho$ term in addition
to                            thermal                           energy
\citep{2000A&A...360.1043D,2010ApJ...719L..28D,2011ApJ...731L..36I,2011ApJ...730...76I}.
For instance, the uncertainty in the  boundary may lead to an order of
magnitude   uncertainty  in   the  value   of  $E_{\rm   bind}$,  and,
consequently,  same  uncertainty  in  the  orbital  separations  of  a
post-CEE binary \citep{2011ApJ...730...76I}.

Another  deficiency of the classic energy  formalism is that,
by  design, the  Equation~(\ref{eq:aform})  implies  that the  kinetic
energy of the  ejected envelope at infinity is  zero (or, in other
words,   is   substantially   smaller    than   the   two   considered
energies).  However,  as has  been  shown  recently  for the  case  of
low-mass giants, if the entire envelope has been successfully ejected,
that envelope  can carry away  between 20\%  and 55\% of  the released
orbital energy,  mainly in the  form of the  kinetic energy, and,  to a
lesser    degree,    in   the    form    of    the   thermal    energy
\citep{2016MNRAS.460.3992N}.

Considering  three  fundamental  energies --  gravitational  potential
energy, thermal energy of the envelope and kinetic energy -- CEEs were
studied  using different  three-dimensional  (3D) hydrodynamic  codes.
Universal evolution of a CEE in 3D simulations is to start a plunge-in
of the  companion, during which  the binary orbit shrinks  strongly on
the timescale comparable to the  initial binary orbital period. By the
end  of the  plunge-in, the  strength of  all frictional  interactions
between  the shrunken  binary and  the inflated  envelope is  strongly
reduced.  The  binary settles into  a slow spiral-in with  a minuscule
orbital                        dissipation                        rate
\citep{2008ApJ...672L..41R,2011MNRAS.411.2277D,2012ApJ...746...74R,2012ApJ...744...52P,2015MNRAS.450L..39N,2016ApJ...816L...9O,2016MNRAS.458..832S}.
Independently  the  type  of  employed  code,  only  partial  envelope
ejections  had  been  obtained.   It  showed  clearly  that  something
essential is missing, and the missing piece is neither the type of the
code, nor  the resolution, but should  be related to physics  that has
not been yet taken into account.

Indeed, there are other, ``non-fundamental'', ways in which the energy
can be generated or lost during a CEE.

One of the sources of energy is  due to accretion on a companion while
it swirls inside  the common envelope.  Energy comes  from the release
of the potential energy of the  accreted material while it reaches the
surface of  the companion, in the  form of heat and  radiation. If the
companion does not  accept all the accreted material,  some energy may
be released back via jets.  Jets inject the kinetic energy back to the
common  envelope,  inflating ``bubbles''  and  helping  to remove  the
common                envelope                 this                way
\citep{2016MNRAS.462..206A,2017MNRAS.465L..54S}.  The total input from
this energy source  depends on the mass of the  companion, on the mass
accretion rate, and  the time during which the  accretion takes place.
To  find the  accretion rate,  a common  way in  the past  was to  use
Bondi-Hoyle-Lyttleton                                     prescription
\citep{1939PCPS...34..405H,1944MNRAS.104..273B,1952MNRAS.112..195B}. It
has been found however that in  3D simulations the accretion rate onto
the companion is significantly  smaller than the Bondi-Hoyle-Lyttleton
prescription would  provide \citep{2012ApJ...746...74R}. On  the other
hand, more recent, albeit simplified,  studies of the accretion during
a   CEE,    have   found    accretion   rates   that    approach   the
Bondi-Hoyle-Lyttleton                                     prescription
\citep{2015ApJ...798L..19M,2015ApJ...803...41M}.  It  is not  clear if
one  of  the  accepted  simplifications, or  the  differences  in  the
considered stellar  models in  3D studies  and in  simplified studies,
have led to the striking difference  in the accretion rates.  The time
on which this  energy can be generated efficiently can  be as small as
the  initial orbital  period, as  this dictates  the timescale  of the
plunge-in.   After  the  plunge,  a   shrunken  binary  clears  out  its
neighborhood and may avoid  continued accretion.  Whatever accretion
rate  would be  eventually  found to  be correct,  the  case when  the
accretion source of energy can become comparable to the binding energy
of the envelope is likely limited  to the case when a companion, while
accreting at its Eddington rate, is spiraling-in to a very large 
donor, $\sim 1000 R_\odot$.

The role of  the magnetic field has also  been contemplated.  Magnetic
fields  were found  to strongly  shape  the outflows  from the  common
envelopes     \citep{2006MNRAS.370.2004N,2007MNRAS.376..599N}.     For
little-bound envelopes of AGB stars,  these magnetic outflows has been
argued to  help to unbind  the entire envelope, although  the complete
ejection was  not directly obtained  in simulations. For  low-mass red
giants,  the presence  of  the  magnetic field  was  determined to  be
dynamically irrelevant for a  common envelope ejection, despite strong
amplification of the magnetic fields \citep{2016MNRAS.462L.121O}.

In some CEEs,  if a non-degenerate companion has  initially failed to
eject the common envelope, and has  to merge with the donor's
core,   the   companion's   material    can   trigger   an   explosive
nucleosynthesis on the outer parts of the core of the evolved donor. This can lead to the
explosive ejection  of not  just the  envelope, but  also of  both the
hydrogen           and           the           helium           layers
\citep{2002PhDT........25I,2010MNRAS.406..840P}.

The energies  listed above  are not  guaranteed to  be present  in all
CEEs.   However, there  is one  source  of energy  which is  naturally
present in all the cases -- the recombination energy.  It is important
that four  phases of  a CEE, qualitatively  different in  the involved
dominant  physical   processes  and  the  timescales,   are  currently
distinguished:  (a) loss  of corotation,  (b) plunge-in  (this is  the
stage  which   is  often  mistaken  for   a  CEE  as  a   whole),  (c)
self-regulating  spiral-in  (this  stage   only  takes  place  if  the
plunge-in did not lead to complete envelope ejection); (d) termination
of  the  self-regulating  phase,   with  either  a  delayed  dynamical
ejection,  or  a  nuclear  ejection,   or  with  a  merger  \citep[for
  qualitative  definitions  of  the  phases,  see][  and  quantitative
  definitions    can     be    found    in    Ivanova     \&    Nandez
  2016]{2011ASPC..447...91I,CEreview}.   Recently, it  was shown  that
during the loss  of corotation, a substantial fraction  of the initial
envelope mass  can be lost before  the CEE enters the  dynamical phase
during which  the energy  formalism is applicable.   The mass  is lost
while  the  the donor  overfills  its  Roche  lobe, but  the  expanded
envelope does  not yet go beyond  $L_2/L_3$ points, and the  phase can
last            for            thousand            of            years
\citep[][]{2015MNRAS.449.4415P,2017MNRAS.465.2092P}.               The
self-regulating   phase   also   could    last   thousand   of   years
\citep{1979A&A....78..167M,2001ASPC..229..239P}.   At  this  timescale
the radiative energy loss from the common envelope surface is becoming
large    enough    to    affect     the    overall    energy    budget
\citep{2002PhDT........25I}.   As it  has appeared,  the recombination
energy plays an  important, while varying, role during  most of stages
of a CEE.

\section{The role of the recombination energy}

As  the common  envelope  expands  and its  material  cools down,  the
ionized  plasma  can  recombine,  releasing binding  energy  which  is
usually referred to as recombination  energy, $\Delta E_{\rm rec}$. We
note that as cooling continues, formation of molecules can take place,
also  releasing energy,  but  here  we will  not  consider the  energy
related to molecule formation.  Recombination energy was suggested to
be helpful for  ejecting outer stellar layers even  before the concept
of    a     CEE,    to    say    nothing     of  the  energy    formalism
\citep[e.g.,][]{1967AJ.....72Q.813L,1967Natur.215..838R}.       Binary
population  synthesis studies  have shown  that the  inclusion of  the
recombination energy in the energy formalism,  as a part of the envelope's
internal  energy,  provides  the  best fits  to  the  observations  of
subdwarf  B stars  \citep{1994MNRAS.270..121H,2002MNRAS.336..449H}. On
the  other hand,  it has  been  argued that  the recombination  energy
cannot help ejecting  the CE, as the most of  the recombination energy
would  leave the  envelope  immediately  in a  form  of radiation,  as
opacity  in the  envelope might  be too  low to  effectively reprocess
energy    released     in    photons    of     specific    wavelengths
\citep{2003MNRAS.343..456S}. We  note that this restriction  is indeed
valid if the optical depth of  the layer where the recombination takes
place is small (it is close  to the photosphere), and the layer itself
is very  thin, so  the released photons  can not be  used to  heat the
envelope material.

The amount of  the recombination energy that is stored  in an envelope
of the  mass $m_{\rm env}$  prior the start  of a CEE,  neglecting the
ionization of  the elements  others than hydrogen  and helium,  can be
evaluated as follows:

\begin{equation}
  \Delta E_{\rm rec} \approx 2.6\times10^{46} \times \frac{m_{\rm env}}{M_\odot}{\rm ergs}\times (X f_{HI}+Yf_{HeII}+1.46Y f_{HeI})
\end {equation}

Here  $X$ is  the  hydrogen  mass fraction,  $Y$  is  the helium  mass
fraction, $f_{HI}$ is  the fraction of hydrogen  that becomes neutral,
$f_{HeI}$  is  the  fraction  of  helium  that  becomes  neutral,  and
$f_{HeII}$  is  the  fraction  of  helium  that  becomes  only  singly
ionized. With a typical value for helium content and assuming complete
recombination from initially completely ionized material, the released
energy can be as high as  $ \Delta E_{\rm rec} \approx 3\times 10^{46}
\times {m_{\rm  env}}/{M_\odot}$ erg.  Comparing this energy  with the
binding energy as  in the Equation~(\ref{eq:aform}), one  can see that
once the radius of the star exceeds $R \gtrsim 127 R_\odot / \lambda$,
a star can be said to have positive total energy even before the start
of the CEE, i.e. it is unbound.  However, first of all, the release of
this energy has to be triggered. Second, this energy should not escape
in a form of radiation, but be reprocessed by the envelope itself.

\subsection{Recombination  during a plunge-in phase}

 As was mentioned above, an unavoidable outcome of 3D simulations of a
 CEE with a  hydrodynamic code that did not  include the recombination
 energy in  the adopted  equation of  state is  to obtain  a plunge-in
 phase, to  eject a  part of  the envelope,  to inflate  the remaining
 bound  envelope well  above the  binary orbit,  and to  start a  slow
 spiral-in, during which the depletion of the binary orbit is becoming
 too   small  to   be   further  treated   by   a  hydrodynamic   code
 \citep{2008ApJ...672L..41R,2011MNRAS.411.2277D,2012ApJ...746...74R,2012ApJ...744...52P,2015MNRAS.450L..39N,2016ApJ...816L...9O,2016MNRAS.458..832S}. The
 primary reason  for this  outcome is the  decoupling of  the shrunken
 binary  orbit   from  the   remaining  inflated  envelope,   as  both
 gravitational   or    viscous   drags   are   becoming    too   small
 \citep{2016MNRAS.462..362I}.   On  the  other hand,  the  very  first
 attempt to include the recombination  energy in the equation of state
 have   resulted   in   the    complete   common   envelope   ejection
 \citep{2015MNRAS.450L..39N}.

This  very  first study,  where  the  common envelope  was  completely
ejected, have  considered the  formation of the  specific double-white
dwarf (DWD) binary WD 1101+364, a well-measured binary system that has
$P_{\rm orb} = 0.145$  d, and a mass ratio of $q =  M_1/M_2 = 0.87 \pm
0.03$, where  $M_1 \simeq 0.31 M_\odot$ and  $M_2 \simeq 0.36 M_\odot$
are  the   masses  of   the  younger   and  older   WDs,  respectively
\citep{1995MNRAS.275..828M}.  DWD binaries are  the best test-site for
CEE as their younger white dwarfs  must have been formed during a CEE,
and their  pre-CEE binary separations  are strongly restricted  by the
well known core-radius relation of  low-mass giants, albeit there is a
fairly    small     dependence    on    the    total     giant    mass
\citep{2006A&A...460..209V}.  Several simulations performed to form WD
1101+364 using the allowed range of the initial binaries and using the
equation of state that did  not include the recombination energy, also
did not unbind the  envelope \citep{2015MNRAS.450L..39N}. The analysis
has  shown that  the binding  energy of  the remaining  bound envelope
could be easily  overcome by the release of  the recombination energy,
if the  recombination energy  release will be  triggered at  the right
time.  This  is exactly  what the  simulations with  the recombination
energy taken into account have shown \citep{2015MNRAS.450L..39N}.

The physics  of the complete  envelope ejection can be  understood via
introduction of the {\it recombination  radius} -- the radius at which
the released  specific recombination energy  is larger than  the local
specific     potential     energy      \citep[for     more     detail,
  see][]{2016MNRAS.462..362I}.     Usually    hydrogen   starts    its
recombination when all helium is already recombined; in this case this
radius   is  $r_{\rm   rec,H}  \approx   105  R_\odot   \times  m_{\rm
  grav}/M_\odot$.   Here   $m_{\rm  grav}$  is  the   mass  within the
recombination radius -- this mass  includes the companion, the core of
the donor, and the mass envelope within $r_{\rm rec,H}$.

During a  CEE, at first,  the frictional forces dissipate  energy from
the binary  orbit and dump the  same energy into the  common envelope.
This  leads to  the  first dynamical  ejection of  a  fraction of  the
envelope, and it is  the ejection that is present in  all the types of
3D simulations,  independent of the  equation of state or  the adopted
method.

If a still bound envelope has  been dynamically expanded beyond of the
envelope's recombination radius, its material  is doomed to be ejected
to infinity via  the recombination outflows on  a dynamical timescale,
leading to  a prompt binary formation  \citep{2016MNRAS.462..362I}. If
the envelope expansion beyond the recombination radius is slow (only a
small  fraction  of   the  envelope  has  been   expanded  beyond  the
recombination radius on a dynamical timescale), a transition to a slow
spiral-in  takes place.   In this  case, recombination  leads to  {\it
  steady} recombination-powered outflows, the  mass lass through these
outflows  can  be slowly  accelerating,  as  $m_{\rm grav}$  decreases
during the continuing mass  loss \citep{2016MNRAS.462..362I}.  We note
that  it has  been  proposed,  but not  yet  verified  against the  3D
outcomes, that in the case when {\it steady} outflows are established,
the  envelope's enthalpy  rather  than the  envelope's thermal  energy
determines  the   outcome  \citep{2011ApJ...731L..36I}.    During  the
transition to a slow-spiral-in, the  remaining bound envelope can also
``fall'' back on  its parabolic trajectory.  Such  a fallback triggers
another partial envelope ejection that acts on a dynamical time and is
presumably   powered   by   the  compression   ionisation   and   then
recombination of the helium layer \citep{2016MNRAS.462..362I}.

Let us  now consider the  efficiency of  the use of  the recombination
energy.  It has  been found that the structure of  ionisation zones in
an expanded common envelope is  drastically different from the same in
unperturbed  stars.  The  zones of  partial ionisation  of helium  and
hydrogen, i.e.  where $f_{HI}$, $f_{HeI}$ and  $f_{HeII}$ are changing
from 0 to 1, are very thick in mass each -- e.g., they can reach $\sim
0.5 M_\odot$ in a low-mass giant.  Hydrogen is still 1\% ionized at an
optical         depth          of         100          or         more
\citep{2015MNRAS.447.2181I,2016MNRAS.462..362I},  although  a  smaller
degree  of  ionisation  can  remain   in  some  cases  closer  to  the
photosphere.   The   recombination  energy   therefore  can   be  well
reprocessed.   Notably,   the  recombination  energy  of   helium  has
absolutely no chance for escape in a  form of radiation and all can be
used for the envelope expansion \citep{2015MNRAS.447.2181I}.

\subsection{Recombination  during a self-regulated spiral-in}

During a self-regulated spiral-in,  the energy transfer throughout the
common envelope, the nuclear energy  generation, and the energy losses
from  envelopes surface  are becoming  important both  for the  energy
budget and for  the thermal structure of the shared  envelope.  At the
same  time, the  orbital period  of  the shrunken  binary is  becoming
substantially  smaller than  the dynamical  timescale of  the inflated
envelope, mandating  a 3D  hydrodynamic code to  switch to  a timestep
which is  extremely small if  compared to  the timescale on  which the
envelope evolves.  As a result  of these complications, no existing 3D
hydrodynamic code is capable of following the self-regulated spiral-in
\citep[we note  that the  first step  towards treating  the convection
  properly has been made recently  by][]{2017_Ohlmann}.  There is a 3D
study that specifically investigated how the plunge-in transits, via a
slow     spiral-in,     into      the     self-regulated     spiral-in
\citep{2016MNRAS.462..362I}.  However,  the simulations had to  end by
the time when the thermal  timescale processes could become important.
Instead  of  3D,  a  common approach  for  studying  a  self-regulated
spiral-in is to use an one-dimensional (1D) stellar code, modifying it
to {\it mimic} CEE conditions,  with a number of simplifications which
could be different  from study to study  \citep[the pioneering studies
  are][and many thereafter]{1978ApJ...222..269T,1979A&A....78..167M}.

In  one  of  the  1D  studies,  it has  been  found  that  during  the
self-regulated  spiral-in, after  the  envelope has  been inflated,  a
delayed  dynamical   instability  initiating  pulsations   of  growing
amplitude takes place \citep{2002PhDT........25I,2002MNRAS.336..449H}.
These growing pulsations might lead to a delayed dynamical ejection of
the envelope, although the ejection itself was not obtained.

\cite{2015MNRAS.447.2181I}  have  explored  1D  CEE  evolution  for  a
low-mass  giant   in  a   systematic  way,  by   introducing  a constant
``heating'' source of the two  types -- uniform heating throughout the
envelope, and a shell-type at the base of the envelope.  As a reaction
to the artificial ``heating'', the  envelope readjusts by expanding to
its new  ``equilibrium'' radius  -- the radius  at which  the inflated
star radiates away the amount of energy that it receives from both the
artificial heating  and the  shell nuclear burning  -- and  is cooling
down.   Double   ionized  helium   starts  its   recombination.   This
recombination is becoming energetically important and can produce an even
higher  rate of  the energy  input  than the  artificial heating.  The
recombination  zones of  once  ionized helium \ and \ hydrogen  \ propagate
inwards in  mass. With high  heating rates and quick  initial envelope
expansion, outer  layers start moving  faster than their  local escape
velocity.  For  moderate heating  rates, the  envelope expands  to its
``equilibrium''         radius         but         is         becoming
unstable.  \cite{2015MNRAS.447.2181I}  determined  that,  due  to  the
expansion of the zones of partial ionisation of hydrogen and helium in
mass, the envelope's pressure-weighted  $\Gamma_1$ becomes less  than
4/3, and almost the entire envelope becomes dynamically unstable.

In further studies, using a similar approach for an artificial heating
source while using a 1D stellar code that includes hydrodynamic terms,
\cite{Clayton_paper1}  found   that  the  heated  envelopes,   if  not
dynamically ejected  at high heating  rates, also become  unstable and
start to experience non-regular pulsations, with the periods between 3
and 20 years.   Some pulsations lead to the ejection  of a fraction of
the  envelope, with  up  to 10\%  of the  envelope  mass escaping  per
ejection  episode.  These  ejections  have a  nature  similar  to  the
shell-triggered  ejections  found  earlier   in  3D  studies  of  slow
spiral-ins \citep{2016MNRAS.462..362I}.

\subsection{Recombination and the outcomes of CEEs}

Two families of the outcomes are expected.

If a CEE has resulted in a prompt binary formation, the ``classical''
$\alpha$ that relates the initial donor and the final orbit, as in the
Equation~(\ref{eq:aform}), can be  as large as one or even  a bit more
that one.   The revised energy  formalism that taken into  account the
energy that  the ejected material  carried away and  the recombination
energy can  be found in \cite{2016MNRAS.460.3992N}.   This revision of
the energy  formalism is based on  the fits of 3D  simulations of CEEs
for  the grid  of  initial  binaries with  low-mass  giant donors  and
low-mass white dwarfs.

If  a   CEE  has   resulted  in   a  self-regulated   spiral-in,  the
``classical''  $\alpha$ is  only about  0.05-0.25, and  the envelope's
material  is lost  in semi-regular  recombination-triggered pulsations
with   an    interval   between   the   ejections    of   3-20   years
\citep{Clayton_paper1}. We  note however  that this value  of $\alpha$
does not  yet take into account  that some material has  been ejected
``dynamically'' before the self-regulated  spiral-in has started, and
more studies are needed.

\subsection{Appearance of the CEEs}
 
All CEEs, including those that end up as mergers, are accompanied by a
dynamical  ejection of  at least  some envelope  material.  As  plasma
expands,  it   cools  down   and  starts  recombination.   Before  the
recombination starts, gas expansion is  adiabatic.  As most of envelope
material has initially  about the same entropy, the  location at which
the recombination  starts is  also similar  for the  ejected material.
Opacities below the  place where gas recombines are  high, while above
they  are low,  at least  until the  cooled gas  can form  dust.  This
recombination front appears as a ``photosphere'' that hides beneath it
the common envelope, for as long  as there is plasma to be recombined.
Once all  material have  recombined, it  reveals the  common envelope.
This model of Wavefront of Cooling and Recombination (WCR) has been proposed
by  \cite{2013Sci...339..433I}.  It  utilizes an  analytical model  of
\cite{1993ApJ...414..712P}, proposed for  hydrogen envelope cooling in
Type II supernovae during the plateau phase.

This WCR model explains naturally curious observational features of the new class of transients, Luminous Red Novae:
\begin{itemize}
\item Large ``apparent'' size and luminosities, plateau phase for the light-curve.
%  \item Plateau phase for the light-curve.
\item ``Red'' color (temperature of the object is about 5000K).
\item Fast decline of luminosities (timescale of the decline is a fraction of the plateau time, and it is much smaller than the inferred dynamical timescale of the object)
\item Spectroscopic velocities, which are few hundreds of km/s, are larger than the expansion rate of the “effective” radius, which are less than a hundred of km/s 
\end{itemize}

\cite{2013Sci...339..433I} have  shown that the range  of the expected
plateau time and  luminosities for stellar mergers  is consistent with
the observed  ranges for  LRNe, and  that the rate  at which  LRNe are
observed can also  be provided by the stellar  mergers.  Some attempts
are made  to fit the observed  light-curves of LRNe. To  fit V1309~Sco
outburst \citep{2011A&A...528A.114T},  \cite{2013Sci...339..433I} have
used Popov's  analytical model, for  which velocities and the  mass of
the  ejecta  were  provided  by 3D  simulations  \citep[detail  of  3D
  simulations are  in][]{2014ApJ...786...39N}. The light-curve  of M31
2015 LRN was  fitted with the merger  of a binary system  in which the
primary star is a $3 -  5.5 M_\odot$ sub-giant branch star with radius
of $30-40 R_\odot$ \citep{2017ApJ...835..282M}.

If  a  CEE  has  entered into  self-regulated  spiral-in,  the  common
envelope object appears as a luminous pulsation variable (note that an
LRN-type   outburst   is   expected   to  precede   this).    On   the
Hertzsprung-Russell   diagram,  the   pulsations   swirl  around   the
equilibrium point,  the position of  which is dictated by  the heating
rate. Depending  on the  heating rate,  that point  can be  located at
$\log_{10} T_{\rm eff} \approx 3.4-3.5$ (while $\log_{10} T_{\rm eff}$
during  the pulsation  can be  changing between  3.2 and  3.7) and  at
$\log_{10} (L/L_\odot) \approx 4.0-4.4$ (while $\log_{10} (L/L_\odot)$
can vary by up to 500  times between the minimum luminosity during the
pulsation,  and  the  maximum  luminosity).  The  pulsations  are  not
symmetric with  time, and the  time that  a heated envelope  spends at
higher than  equilibrium luminosity is  much smaller than the  time it
takes for the  star to be ``re-heated'' back to  its equilibrium value
\citep[for examples of light-curves, see][]{Clayton_paper1}.

However, if  a CEE had  neither resulted in  a clean merger,  nor had
entered in  a self-regulated  spiral-in, the  observational signatures
are less understood. While the first dynamical ejection can provide an
LRN-type outburst,  further outflows  take places when  some initially
available  recombination  energy  has  been processed  to  unbind  the
envelope. This may  change the observed luminosities,  presence of the
plateau, and  the timescale of  the outbursts.  No  self-consistent 3D
modeling of a CEE leading to a binary formation inclusive of radiative
energy loss have been done yet, and is the important subject of future
studies.


\begin{thebibliography}{} 

\bibitem[Akashi \& Soker(2016)]{2016MNRAS.462..206A} Akashi, M., \& Soker, N.\ 2016,  \textit{MNRAS}, 462, 206 

\bibitem[Bondi \& Hoyle(1944)]{1944MNRAS.104..273B} Bondi, H., \& Hoyle, F.\ 1944,  \textit{MNRAS}, 104, 273 

\bibitem[Bondi(1952)]{1952MNRAS.112..195B} Bondi, H.\ 1952, \textit{MNRAS}, 112, 195 
  
\bibitem[Clayton et al.(2017)]{Clayton_paper1} Clayton, M., Podsiadlowski, P., Ivanova, N., \& Justham, S.\ 2017, \textit{MNRAS}, accepted 

\bibitem[Deloye \& Taam(2010)]{2010ApJ...719L..28D} Deloye, C.~J., \& Taam, R.~E.\ 2010, \textit{ApJ} (Letters), 719, L28 

\bibitem[De Marco et al.(2011)]{2011MNRAS.411.2277D} De Marco, O., Passy, J.-C., Moe, M., et al.\ 2011, \textit{MNRAS}, 411, 2277 

\bibitem[Dewi \& Tauris(2000)]{2000A&A...360.1043D} Dewi, J.~D.~M., \& Tauris, T.~M.\ 2000, \textit{A\&A},360, 1043 

\bibitem[Han et al.(1994)]{1994MNRAS.270..121H} Han, Z., Podsiadlowski, P., \& Eggleton, P.~P.\ 1994, \textit{MNRAS}, 270, 121 
      
\bibitem[Han et al.(2002)]{2002MNRAS.336..449H} Han, Z.,  {\it et al.} \ 2002, \textit{MNRAS}, 336, 449 

\bibitem[Hoyle \& Lyttleton(1939)]{1939PCPS...34..405H} Hoyle, F., \& Lyttleton, R.~A.\ 1939, Proceedings of the Cambridge Philosophical Society, 34, 405 
      
\bibitem[Ivanova(2002)]{2002PhDT........25I} Ivanova, N.\ 2002, Ph.D.~Thesis,  
      
\bibitem[Ivanova(2011a)]{2011ApJ...730...76I} Ivanova, N.\ 2011a, \textit{ApJ}, 730, 76 

\bibitem[Ivanova(2011b)]{2011ASPC..447...91I} Ivanova, N.\ 2011b, Evolution of Compact Binaries, 447, 91 

\bibitem[Ivanova \& Chaichenets(2011)]{2011ApJ...731L..36I} Ivanova, N., \& Chaichenets, S.\ 2011, \textit{ApJ} (Letters), 731, L36 

\bibitem[Ivanova et al.(2013)]{2013Sci...339..433I} Ivanova, N., Justham, S., Avendano Nandez, J.~L., \& Lombardi, J.~C.\ 2013, \textit{Science}, 339, 433 
  
\bibitem[Ivanova et al.(2013)]{CEreview} Ivanova, N., Justham, S., Chen, X., et al.\ 2013, \textit{A\&ARv}, 21, 59 

\bibitem[Ivanova et al.(2015)]{2015MNRAS.447.2181I} Ivanova, N., Justham, S., \& Podsiadlowski, P.\ 2015, \textit{MNRAS}, 447, 2181 
    
\bibitem[Ivanova \& Nandez(2016)]{2016MNRAS.462..362I} Ivanova, N., \& Nandez, J.~L.~A.\ 2016, \textit{MNRAS}, 462, 362 
      
\bibitem[Livio \& Soker(1988)]{1988ApJ...329..764L} Livio, M., \& Soker, N.\ 1988, \textit{ApJ}, 329, 764 

\bibitem[Lucy(1967)]{1967AJ.....72Q.813L} Lucy, L.~B.\ 1967, \textit{AJ}, 72, 813 
    
\bibitem[MacLeod \& Ramirez-Ruiz(2015a)]{2015ApJ...798L..19M} MacLeod, M., \& Ramirez-Ruiz, E.\ 2015a, \textit{ApJ} (Letters), 798, L19 

\bibitem[MacLeod \& Ramirez-Ruiz(2015b)]{2015ApJ...803...41M} MacLeod, M., \& Ramirez-Ruiz, E.\ 2015b, \textit{ApJ}, 803, 41 

\bibitem[MacLeod et al.(2017)]{2017ApJ...835..282M} MacLeod, M., Macias, P., Ramirez-Ruiz, E., et al.\ 2017, \textit{ApJ}, 835, 282 
  
\bibitem[Marsh et al.(1995)]{1995MNRAS.275..828M} Marsh, T.~R., Dhillon, V.~S., \& Duck, S.~R.\ 1995, \textit{MNRAS}, 275, 828 

\bibitem[Meyer \& Meyer-Hofmeister(1979)]{1979A&A....78..167M} Meyer, F., \& Meyer-Hofmeister, E.\ 1979, \textit{A\&A},78, 167 

\bibitem[Nandez et al.(2014)]{2014ApJ...786...39N} Nandez, J.~L.~A., Ivanova, N., \& Lombardi, J.~C., Jr.\ 2014, \textit{ApJ}, 786, 39 
  
\bibitem[Nandez et al.(2015)]{2015MNRAS.450L..39N} Nandez, J.~L.~A., Ivanova, N., \& Lombardi, J.~C.\ 2015, \textit{MNRAS}, 450, L39 
    
\bibitem[Nandez \& Ivanova(2016)]{2016MNRAS.460.3992N} Nandez, J.~L.~A., \& Ivanova, N.\ 2016, \textit{MNRAS}, 460, 3992 

\bibitem[Nordhaus \& Blackman(2006)]{2006MNRAS.370.2004N} Nordhaus, J., \& Blackman, E.~G.\ 2006, \textit{MNRAS}, 370, 2004 
  
\bibitem[Nordhaus et al.(2007)]{2007MNRAS.376..599N} Nordhaus, J., Blackman, E.~G., \& Frank, A.\ 2007, \textit{MNRAS}, 376, 599 

\bibitem[Ohlmann et al.(2016a)]{2016ApJ...816L...9O} Ohlmann, S.~T., R{\"o}pke, F.~K., Pakmor, R., \& Springel, V.\ 2016a, \textit{ApJ} (Letters), 816, L9 

\bibitem[Ohlmann et al.(2016b)]{2016MNRAS.462L.121O} Ohlmann, S.~T., {\it et al.}  \ 2016b, \textit{MNRAS}, 462, L121 

\bibitem[Ohlmann et al.(2017)]{2017_Ohlmann} Ohlmann, S.~T., Roepke, F.~K., Pakmor, R., \& Springel, V.\ 2017, \textit{A\&A}

\bibitem[Passy et al.(2012)]{2012ApJ...744...52P} Passy, J.-C., De Marco, O., Fryer, C.~L., et al.\ 2012, \textit{ApJ}, 744, 52 

\bibitem[Pavlovskii \& Ivanova(2015)]{2015MNRAS.449.4415P} Pavlovskii, K., \& Ivanova, N.\ 2015, \textit{MNRAS}, 449, 4415 
  
\bibitem[Pavlovskii et al.(2017)]{2017MNRAS.465.2092P} Pavlovskii, K., Ivanova, N., Belczynski, K., \& Van, K.~X.\ 2017, \textit{MNRAS}, 465, 2092 
  
\bibitem[Podsiadlowski(2001)]{2001ASPC..229..239P} Podsiadlowski, P.\ 2001, Evolution of Binary and Multiple Star Systems, 229, 239 
  
\bibitem[Podsiadlowski et al.(2010)]{2010MNRAS.406..840P} Podsiadlowski, P., Ivanova, N., Justham, S., \& Rappaport, S.\ 2010, \textit{MNRAS}, 406, 840 

\bibitem[Popov(1993)]{1993ApJ...414..712P} Popov, D.~V.\ 1993, \textit{ApJ}, 414, 712 
  
\bibitem[Ricker \& Taam(2008)]{2008ApJ...672L..41R} Ricker, P.~M., \& Taam, R.~E.\ 2008, \textit{ApJ} (Letters), 672, L41 

\bibitem[Ricker \& Taam(2012)]{2012ApJ...746...74R} Ricker, P.~M., \& Taam, R.~E.\ 2012, \textit{ApJ}, 746, 74 

\bibitem[Roxburgh(1967)]{1967Natur.215..838R} Roxburgh, I.~W.\ 1967, \textit{Nature}, 215, 838 
  
\bibitem[Shiber et al.(2017)]{2017MNRAS.465L..54S} Shiber, S., Kashi, A., \& Soker, N.\ 2017, \textit{MNRAS}, 465, L54 

\bibitem[Soker \& Harpaz(2003)]{2003MNRAS.343..456S} Soker, N., \& Harpaz, A.\ 2003, \textit{MNRAS}, 343, 456 
  
\bibitem[Staff et al.(2016)]{2016MNRAS.458..832S} Staff, J.~E., De Marco, O., Wood, P., Galaviz, P., \& Passy, J.-C.\ 2016, \textit{MNRAS}, 458, 832 

\bibitem[Taam et al.(1978)]{1978ApJ...222..269T} Taam, R.~E., Bodenheimer, P., \& Ostriker, J.~P.\ 1978, \textit{ApJ}, 222, 269 

\bibitem[Tylenda et al.(2011)]{2011A&A...528A.114T} Tylenda, R., Hajduk, M., Kami{\'n}ski, T., et al.\ 2011, \textit{A\&A},528, A114 

\bibitem[van der Sluys et al.(2006)]{2006A&A...460..209V} van der Sluys, M.~V., Verbunt, F., \& Pols, O.~R.\ 2006, \textit{A\&A},460, 209 
  
\bibitem[Webbink(1984)]{1984ApJ...277..355W} Webbink, R.~F.\ 1984, \textit{ApJ}, 277, 355 

  
\end{thebibliography}
\end{document}